\documentstyle[10pt,epsf,epsfig,dp_delphititle]{dp_delphi}
\makeindex
\def\DpPaperGroup{EP}
\def\DpPaperRef{2000-124}
\def\DpDate{28 August 2000}
\def\DpAuthors{DELPHI Collaboration}
\def\DpSubmit{(Submitted to Physics Letter B)}
\def\DpTitle{{Search for Spontaneous  $R$-parity violation 
at $\sqrt{s} =$ 183\,GeV and 189\,GeV}}
\def\DpComment{ }
\def\DpEMail{}

\begin{document}
\makeatletter
\newcount\@tempcntc
\def\@citex[#1]#2{\if@filesw\immediate\write\@auxout{\string\citation{#2}}\fi
  \@tempcnta\z@\@tempcntb\m@ne\def\@citea{}\@cite{\@for\@citeb:=#2\do
    {\@ifundefined
       {b@\@citeb}{\@citeo\@tempcntb\m@ne\@citea\def\@citea{,}{\bf ?}\@warning
       {Citation `\@citeb' on page \thepage \space undefined}}%
    {\setbox\z@\hbox{\global\@tempcntc0\csname b@\@citeb\endcsname\relax}%
     \ifnum\@tempcntc=\z@ \@citeo\@tempcntb\m@ne
       \@citea\def\@citea{,}\hbox{\csname b@\@citeb\endcsname}%
     \else
      \advance\@tempcntb\@ne
      \ifnum\@tempcntb=\@tempcntc
      \else\advance\@tempcntb\m@ne\@citeo
      \@tempcnta\@tempcntc\@tempcntb\@tempcntc\fi\fi}}\@citeo}{#1}}
\def\@citeo{\ifnum\@tempcnta>\@tempcntb\else\@citea\def\@citea{,}%
  \ifnum\@tempcnta=\@tempcntb\the\@tempcnta\else
   {\advance\@tempcnta\@ne\ifnum\@tempcnta=\@tempcntb \else \def\@citea{--}\fi
    \advance\@tempcnta\m@ne\the\@tempcnta\@citea\the\@tempcntb}\fi\fi}
 
\makeatother
\begin{titlepage}
\pagenumbering{roman}
\CERNpreprint{\DpPaperGroup}{\DpPaperRef} 
\date{{\small\DpDate}} 
\title{\DpTitle} 
\address{\DpAuthors} 
\begin{shortabs} 
\noindent
%
\noindent
Searches for spontaneous $R$-parity violating signals
at $\sqrt{s}=183$\,GeV and \mbox{$\sqrt{s}=189$\,GeV} have been performed using the 
1997 and 1998 DELPHI data, under the assumption of $R$-parity breaking in
the third lepton family. The expected topology for the decay of a pair of
charginos into two acoplanar taus plus missing energy was investigated and
no evidence for a signal was found. The results were used to derive a limit on
the chargino mass and to constrain the allowed domains of the MSSM parameter space.

\end{shortabs}
\vfill
\begin{center}
\DpSubmit \ \\ 
\DpComment \ \\
\DpEMail \ \\
\end{center}
\vfill
\clearpage
\headsep 10.0pt
\addtolength{\textheight}{10mm}
\addtolength{\footskip}{-5mm}
\begingroup
%
\newcommand{\DpName}[2]{\hbox{#1$^{\ref{#2}}$},\hfill}
\newcommand{\DpNameTwo}[3]{\hbox{#1$^{\ref{#2},\ref{#3}}$},\hfill}
\newcommand{\DpNameThree}[4]{\hbox{#1$^{\ref{#2},\ref{#3},\ref{#4}}$},\hfill}
\newskip\Bigfill \Bigfill = 0pt plus 1000fill
\newcommand{\DpNameLast}[2]{\hbox{#1$^{\ref{#2}}$}\hspace{\Bigfill}}
%
\footnotesize
\noindent
\DpName{P.Abreu}{LIP}
\DpName{W.Adam}{VIENNA}
\DpName{T.Adye}{RAL}
\DpName{P.Adzic}{DEMOKRITOS}
\DpName{Z.Albrecht}{KARLSRUHE}
\DpName{T.Alderweireld}{AIM}
\DpName{G.D.Alekseev}{JINR}
\DpName{R.Alemany}{CERN}
\DpName{T.Allmendinger}{KARLSRUHE}
\DpName{P.P.Allport}{LIVERPOOL}
% ...
\normalsize
\endgroup
\titlefoot{Department of Physics and Astronomy, Iowa State
     University, Ames IA 50011-3160, USA
    \label{AMES}}
\titlefoot{Physics Department, Univ. Instelling Antwerpen,
     Universiteitsplein 1, B-2610 Antwerpen, Belgium \\
     \indent~~and IIHE, ULB-VUB,
     Pleinlaan 2, B-1050 Brussels, Belgium \\
     \indent~~and Facult\'e des Sciences,
     Univ. de l'Etat Mons, Av. Maistriau 19, B-7000 Mons, Belgium
    \label{AIM}}
\titlefoot{Physics Laboratory, University of Athens, Solonos Str.
     104, GR-10680 Athens, Greece
    \label{ATHENS}}
\titlefoot{Department of Physics, University of Bergen,
     All\'egaten 55, NO-5007 Bergen, Norway
    \label{BERGEN}}
\titlefoot{Dipartimento di Fisica, Universit\`a di Bologna and INFN,
     Via Irnerio 46, IT-40126 Bologna, Italy
    \label{BOLOGNA}}
\titlefoot{Centro Brasileiro de Pesquisas F\'{\i}sicas, rua Xavier Sigaud 150,
     BR-22290 Rio de Janeiro, Brazil \\
     \indent~~and Depto. de F\'{\i}sica, Pont. Univ. Cat\'olica,
     C.P. 38071 BR-22453 Rio de Janeiro, Brazil \\
     \indent~~and Inst. de F\'{\i}sica, Univ. Estadual do Rio de Janeiro,
     rua S\~{a}o Francisco Xavier 524, Rio de Janeiro, Brazil
    \label{BRASIL}}
\titlefoot{Comenius University, Faculty of Mathematics and Physics,
     Mlynska Dolina, SK-84215 Bratislava, Slovakia
    \label{BRATISLAVA}}
\titlefoot{Coll\`ege de France, Lab. de Physique Corpusculaire, IN2P3-CNRS,
     FR-75231 Paris Cedex 05, France
    \label{CDF}}
\titlefoot{CERN, CH-1211 Geneva 23, Switzerland
    \label{CERN}}
\titlefoot{Institut de Recherches Subatomiques, IN2P3 - CNRS/ULP - BP20,
     FR-67037 Strasbourg Cedex, France
    \label{CRN}}
\titlefoot{Now at DESY-Zeuthen, Platanenallee 6, D-15735 Zeuthen, Germany
    \label{DESY}}
\titlefoot{Institute of Nuclear Physics, N.C.S.R. Demokritos,
     P.O. Box 60228, GR-15310 Athens, Greece
    \label{DEMOKRITOS}}
\titlefoot{FZU, Inst. of Phys. of the C.A.S. High Energy Physics Division,
     Na Slovance 2, CZ-180 40, Praha 8, Czech Republic
    \label{FZU}}
\titlefoot{Currently at DPNC,
     University of Geneva,
     Quai Ernest-Ansermet 24, CH-1211, Geneva, Switzerland
    \label{GENEVA}}
\titlefoot{Dipartimento di Fisica, Universit\`a di Genova and INFN,
     Via Dodecaneso 33, IT-16146 Genova, Italy
    \label{GENOVA}}
\titlefoot{Institut des Sciences Nucl\'eaires, IN2P3-CNRS, Universit\'e
     de Grenoble 1, FR-38026 Grenoble Cedex, France
    \label{GRENOBLE}}
\titlefoot{Helsinki Institute of Physics, HIP,
     P.O. Box 9, FI-00014 Helsinki, Finland
    \label{HELSINKI}}
\titlefoot{Joint Institute for Nuclear Research, Dubna, Head Post
     Office, P.O. Box 79, RU-101 000 Moscow, Russian Federation
    \label{JINR}}
\titlefoot{Institut f\"ur Experimentelle Kernphysik,
     Universit\"at Karlsruhe, Postfach 6980, DE-76128 Karlsruhe,
     Germany
    \label{KARLSRUHE}}
\titlefoot{Institute of Nuclear Physics and University of Mining and Metalurgy,
     Ul. Kawiory 26a, PL-30055 Krakow, Poland
    \label{KRAKOW}}
\titlefoot{Universit\'e de Paris-Sud, Lab. de l'Acc\'el\'erateur
     Lin\'eaire, IN2P3-CNRS, B\^{a}t. 200, FR-91405 Orsay Cedex, France
    \label{LAL}}
\titlefoot{LIP, IST, FCUL - Av. Elias Garcia, 14-$1^{o}$,
     PT-1000 Lisboa Codex, Portugal
    \label{LIP}}
\titlefoot{Department of Physics, University of Liverpool, P.O.
     Box 147, Liverpool L69 3BX, UK
    \label{LIVERPOOL}}
\titlefoot{LPNHE, IN2P3-CNRS, Univ.~Paris VI et VII, Tour 33 (RdC),
     4 place Jussieu, FR-75252 Paris Cedex 05, France
    \label{LPNHE}}
\titlefoot{Department of Physics, University of Lund,
     S\"olvegatan 14, SE-223 63 Lund, Sweden
    \label{LUND}}
\titlefoot{Universit\'e Claude Bernard de Lyon, IPNL, IN2P3-CNRS,
     FR-69622 Villeurbanne Cedex, France
    \label{LYON}}
\titlefoot{Univ. d'Aix - Marseille II - CPP, IN2P3-CNRS,
     FR-13288 Marseille Cedex 09, France
    \label{MARSEILLE}}
\titlefoot{Dipartimento di Fisica, Universit\`a di Milano and INFN-MILANO,
     Via Celoria 16, IT-20133 Milan, Italy
    \label{MILANO}}
\titlefoot{Dipartimento di Fisica, Univ. di Milano-Bicocca and
     INFN-MILANO, Piazza delle Scienze 2, IT-20126 Milan, Italy
    \label{MILANO2}}
\titlefoot{IPNP of MFF, Charles Univ., Areal MFF,
     V Holesovickach 2, CZ-180 00, Praha 8, Czech Republic
    \label{NC}}
\titlefoot{NIKHEF, Postbus 41882, NL-1009 DB
     Amsterdam, The Netherlands
    \label{NIKHEF}}
\titlefoot{National Technical University, Physics Department,
     Zografou Campus, GR-15773 Athens, Greece
    \label{NTU-ATHENS}}
\titlefoot{Physics Department, University of Oslo, Blindern,
     NO-1000 Oslo 3, Norway
    \label{OSLO}}
\titlefoot{Dpto. Fisica, Univ. Oviedo, Avda. Calvo Sotelo
     s/n, ES-33007 Oviedo, Spain
    \label{OVIEDO}}
\titlefoot{Department of Physics, University of Oxford,
     Keble Road, Oxford OX1 3RH, UK
    \label{OXFORD}}
\titlefoot{Dipartimento di Fisica, Universit\`a di Padova and
     INFN, Via Marzolo 8, IT-35131 Padua, Italy
    \label{PADOVA}}
\titlefoot{Rutherford Appleton Laboratory, Chilton, Didcot
     OX11 OQX, UK
    \label{RAL}}
\titlefoot{Dipartimento di Fisica, Universit\`a di Roma II and
     INFN, Tor Vergata, IT-00173 Rome, Italy
    \label{ROMA2}}
\titlefoot{Dipartimento di Fisica, Universit\`a di Roma III and
     INFN, Via della Vasca Navale 84, IT-00146 Rome, Italy
    \label{ROMA3}}
\titlefoot{DAPNIA/Service de Physique des Particules,
     CEA-Saclay, FR-91191 Gif-sur-Yvette Cedex, France
    \label{SACLAY}}
\titlefoot{Instituto de Fisica de Cantabria (CSIC-UC), Avda.
     los Castros s/n, ES-39006 Santander, Spain
    \label{SANTANDER}}
\titlefoot{Dipartimento di Fisica, Universit\`a degli Studi di Roma
     La Sapienza, Piazzale Aldo Moro 2, IT-00185 Rome, Italy
    \label{SAPIENZA}}
\titlefoot{Inst. for High Energy Physics, Serpukov
     P.O. Box 35, Protvino, (Moscow Region), Russian Federation
    \label{SERPUKHOV}}
\titlefoot{J. Stefan Institute, Jamova 39, SI-1000 Ljubljana, Slovenia
     and Laboratory for Astroparticle Physics,\\
     \indent~~Nova Gorica Polytechnic, Kostanjeviska 16a, SI-5000 Nova Gorica, Slovenia, \\
     \indent~~and Department of Physics, University of Ljubljana,
     SI-1000 Ljubljana, Slovenia
    \label{SLOVENIJA}}
\titlefoot{Fysikum, Stockholm University,
     Box 6730, SE-113 85 Stockholm, Sweden
    \label{STOCKHOLM}}
\titlefoot{Dipartimento di Fisica Sperimentale, Universit\`a di
     Torino and INFN, Via P. Giuria 1, IT-10125 Turin, Italy
    \label{TORINO}}
\titlefoot{Dipartimento di Fisica, Universit\`a di Trieste and
     INFN, Via A. Valerio 2, IT-34127 Trieste, Italy \\
     \indent~~and Istituto di Fisica, Universit\`a di Udine,
     IT-33100 Udine, Italy
    \label{TU}}
\titlefoot{Univ. Federal do Rio de Janeiro, C.P. 68528
     Cidade Univ., Ilha do Fund\~ao
     BR-21945-970 Rio de Janeiro, Brazil
    \label{UFRJ}}
\titlefoot{Department of Radiation Sciences, University of
     Uppsala, P.O. Box 535, SE-751 21 Uppsala, Sweden
    \label{UPPSALA}}
\titlefoot{IFIC, Valencia-CSIC, and D.F.A.M.N., U. de Valencia,
     Avda. Dr. Moliner 50, ES-46100 Burjassot (Valencia), Spain
    \label{VALENCIA}}
\titlefoot{Institut f\"ur Hochenergiephysik, \"Osterr. Akad.
     d. Wissensch., Nikolsdorfergasse 18, AT-1050 Vienna, Austria
    \label{VIENNA}}
\titlefoot{Inst. Nuclear Studies and University of Warsaw, Ul.
     Hoza 69, PL-00681 Warsaw, Poland
    \label{WARSZAWA}}
\titlefoot{Fachbereich Physik, University of Wuppertal, Postfach
     100 127, DE-42097 Wuppertal, Germany
    \label{WUPPERTAL}}
\addtolength{\textheight}{-10mm}
\addtolength{\footskip}{5mm}
\clearpage
\headsep 30.0pt
\end{titlepage}
%
\pagenumbering{arabic} 
\setcounter{footnote}{0} %
\large
%
%
%
\section{Introduction}

$R$-parity is a discrete symmetry assigned as $R_p=(-1)^{3B+L+2S}$, where 
$B$ is the baryon number, $L$ is the lepton number and $S$ is the particle spin.
In the Minimal Supersymmetric Standard Model (MSSM) the $R$-parity symmetry 
is assumed to be conserved~\cite{mssm}. 
Under this assumption the supersymmetric particles must be produced in 
pairs, every SUSY particle decays into another SUSY particle and the 
lightest of them is absolutely stable. These features underly most of 
the experimental searches for supersymmetric states.

One alternative supersymmetric scenario is to consider the $R$-parity as an 
exact Lagrangian symmetry, broken spontaneously through the Higgs 
mechanism~\cite{rp}. This may take place via non-zero vacuum expectation 
values (VEVs) for scalar neutrinos, such as for the scalar tau-neutrinos
\begin{equation} 
\begin{array}{cc}
v_R = \langle\tilde{\nu}_{R\tau}\rangle~;~  &
v_L = \langle\tilde{\nu}_{L\tau}\rangle~. 
\end{array}
\end{equation} 
In this case there are two main scenarios depending on whether the lepton 
number is a 
gauge symmetry or not~\cite{rp1,rp-model,fernando,rp2,rp3}. In the absence 
of an additional 
gauge symmetry, it leads to the existence of a physical massless 
Nambu-Goldstone boson, called the Majoron (J)~\cite{rp-model}. 
In this context the Majoron remains massless and therefore stable provided that there 
are no explicit $R$-parity violating terms.

\subsection{Spontaneous R-Parity violation}
In the present work we consider the simplest version of the $R$-parity 
spontaneous violation model described in Ref.~\cite{rp-model,fernando}. In 
this model the Lagrangian is specified by the superpotential 
\begin{equation}
W = W_1 + h_\nu\nu^c L{\mathrm H_u} + h\Phi\nu^c{\mathrm S} + h.c.
\end{equation}
that conserves the total lepton number and $R$-parity. The first part of this 
equation contains the basic MSSM superpotential terms, including an isosinglet 
scalar $\Phi$ with a linear superpotential coupling, written as: 
\begin{equation}
W_1= h_uQu^c{\mathrm H_u} + h_dQd^c{\mathrm H_d} + 
h_ee^cL{\mathrm H_d} + (h_0{\mathrm H_uH_d}-\mu'^2)\Phi~. 
\end{equation}
The couplings 
$h_u$, $h_d$, $h_e$, $h_\nu$, $h_0$, $h$ are described by arbitrary 
matrices in the generation space and explicitly break flavour conservation. 
The additional chiral superfields $\nu^c$, $S$~\cite{Smatriz} and 
$\Phi$~\cite{Phimatrices} are singlets under $SU(2)\bigotimes U(1)$ and carry a 
conserved lepton number assigned as -1, 1 and 0, respectively. These 
superfields 
may induce the spontaneous violation of $R$-parity,
given by the imaginary part of: 
\begin{equation}
\frac{v_L^2}{Vv^2}(v_u{\mathrm H_u}-v_d{\mathrm H_d}) + \frac{v_L}{V}\tilde\nu_\tau 
-\frac{v_R}{V}\tilde\nu_\tau^c + \frac{v_S}{V}\tilde {\mathrm S_\tau}~,
\end{equation}
leading to an $R$-odd Majoron. The isosinglet VEVs 
$v_R=\langle\tilde{\nu}_{R\tau}\rangle$ and 
$v_S = \langle\tilde S_\tau\rangle$, with $V=\sqrt{v_R^2+v_S^2}$, 
characterise the $R$-parity breaking and the isodoublet VEVs 
$v_u=\langle H_u\rangle$, 
$v_d=\langle H_d\rangle$ and $v_L=\langle\tilde{\nu}_{L\tau}\rangle$ induce the electroweak 
breaking and generate the fermion masses. For theoretical reasons the 
$R$-parity 
breaking was introduced only in the third family, since the largest Yukawa 
couplings are those of the third generation.
In that case the $R$-parity breaking is effectively parameterised by a 
bilinear superpotential term given by:
\begin{equation}
\epsilon_i\equiv h_{\nu_{i3}}v_{R3}~. 
\end{equation}
This effective parameter leads to the $R$-parity violating gauge couplings and 
contributes to the mixing between the charged (neutral) leptons and the 
charginos (neutralinos), as can be seen from the fermion mass matrices in 
Ref.~\cite{mass}.

By construction, neutrinos are massless at the Lagrangian level but get 
mass from the mixing with neutralinos~\cite{rp2,mass}. As a result, 
all $R$-parity violating observables are directly correlated to the 
$\tau$ neutrino mass:
\begin{equation}
m_{\nu_\tau}\sim \frac{\xi\epsilon^2}{m_{\tilde{\chi}}}~,
\end{equation}
where $m_{\tilde{\chi}}$ is the neutralino mass, $\epsilon$ is the $R$-parity violation parameter and $\xi$ is an effective parameter~\cite{neutrino} given as a function of M$_2$, $\mu$ and 
$\tan\beta$. 

\subsection{Chargino Decay Modes}
\begin{figure}
\begin{center}
\mbox{\epsfysize=13.cm\epsffile{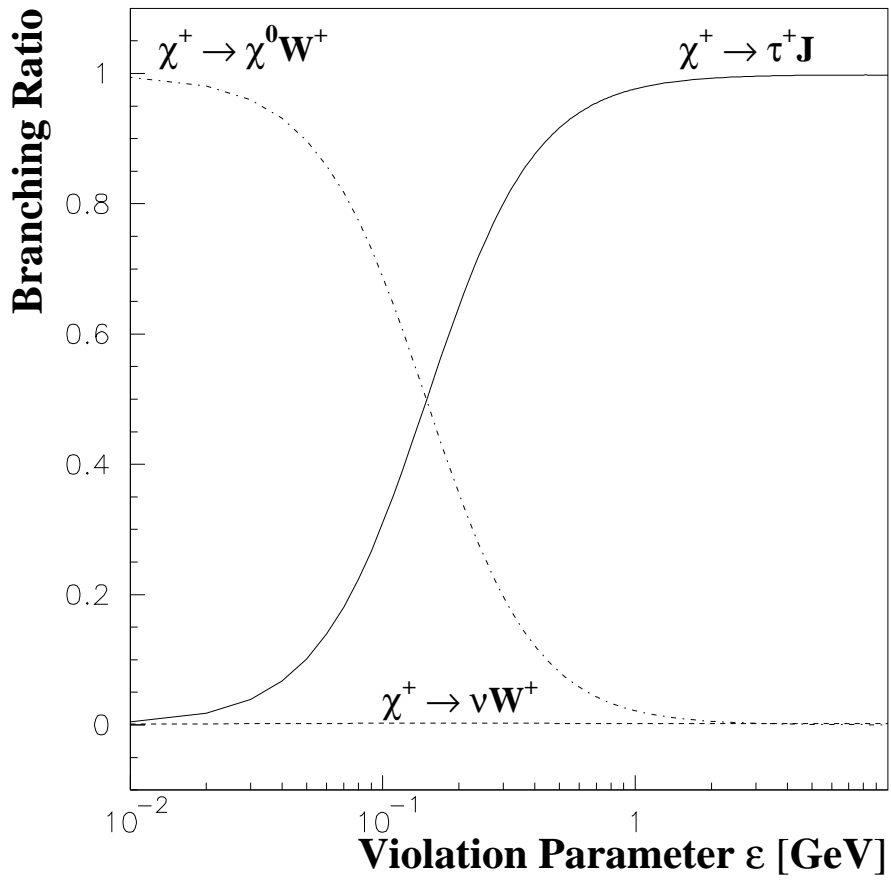}}
\protect\caption{Chargino decay branching ratios as a function of the 
effective $R$-parity violation parameter $\epsilon$ for $\tan\beta=2$, 
$\mu=100$\,GeV/c$^2$ and M$_2=400$\,GeV/c$^2$.}
\protect\label{branch}
\end{center}
\end{figure}

At LEP2 the chargino can be pair produced from $e^+e^-$ via exchange of  
$\gamma$, Z or $\tilde\nu$. In the present analysis it is assumed that all 
sfermions are sufficiently heavy ($\rm{M}_{\tilde{\nu}}\geq 300$\,GeV/c$^2$)
not to influence the chargino production or decay. Therefore, 
only the $\gamma$ and Z $s$-channels contribute to the chargino cross-section. 
In the spontaneous $R$-parity violation model with $R$-parity 
breaking in the third generation, the lightest chargino 
($\tilde{\chi}^\pm$) can undergo a two-body decay mode with a Majoron (J) 
in the final state 
\begin{equation}
\label{twobody}
\tilde{\chi}^\pm \rightarrow \tau^\pm J \\
\end{equation}
in addition to the ``conventional" chargino channels  
\begin{equation}
\label{rpv}
\tilde{\chi}^\pm \rightarrow \nu_\tau \rm{W}^\pm \rightarrow 
                             \nu_\tau q\bar{q'},~\nu_\tau l^\pm_i\nu_i 
\end{equation}
and 
\begin{equation}
\label{rpc}
\tilde{\chi}^\pm \rightarrow \tilde{\chi}^0 \rm{W}^\pm \rightarrow
                       \tilde{\chi}^0 q\bar{q'},~\tilde{\chi}^0 l^\pm_i\nu_i ~.
\end{equation}
Both the two-body decay (\ref{twobody}) and the decay with a neutralino in the 
final state (\ref{rpc}) are $R$-parity conserving, 
while in equation (\ref{rpv}) the chargino decays through an $R$-parity violating vertex.
The decay branching ratios depend strongly on the effective $R$-parity  
violation parameter ($\epsilon$), as can be observed in figure~\ref{branch}.
Note that in a large range of $\epsilon$ the new two-body decay 
mode is the dominant channel and, since it is $R$-parity conserving, it can be 
large.  

\subsection{Parameter Values}
All the results discussed in the following sections were achieved by 
assuming that the chargino decays mainly via the new two body decay mode, 
described in equation (\ref{twobody}). 
As was already mentioned, all sfermions are considered to be  
sufficiently heavy (M$_{\tilde{\nu}}\geq 300$\,GeV/c$^2$) 
not to influence the chargino production or decay. 
Typical ranges of values for the SUSY parameters 
$\mu\equiv h_0\langle\Phi\rangle$ and M$_2$ are assumed: 
\begin{equation}
-200\,\rm{GeV}/c^2 \le \mu \le 200\,\rm{GeV}/c^2 
\end{equation}
\begin{equation}
40\,\rm{GeV}/c^2 \le \mathrm{M}_2 \le 400\,\rm{GeV}/c^2~, 
\end{equation}
which can be covered by the chargino production at LEP.
Also assumed are the GUT relation M$_1$/M$_2=5/3\tan^2\theta_W$ and  
that $\tan\beta~(={v_u}/{v_d})$ lies in the range
\begin{equation}
2 \le \tan\beta \le 40 ~. 
\end{equation}


\section{Detector Description}

The following is a summary of the properties of the DELPHI 
detector~\cite{delphi} relevant to this analysis.
Charged particle tracks were reconstructed in the 1.2 T solenoidal 
magnetic field by a system of cylindrical tracking detectors.
 These were the microVertex Detector (VD),
the Inner Detector (ID), the Time Projection Chamber
(TPC), and the Outer Detector (OD). In addition,
two planes of drift chambers aligned
perpendicular to the  beam axis (Forward Chambers A and B) tracked
particles in the forward and backward directions, covering polar
angles $11^\circ<\theta<33^\circ$ and $147^\circ<\theta<169^\circ$
with respect to the beam ($z$) direction.

The VD consisted of three cylindrical layers of 
silicon detectors, at radii 6.3~cm, 9.0~cm and 11.0~cm.
All three layers measured 
coordinates in the plane transverse to the beam.
The closest (6.3~cm) and the outer (11.0~cm) layers
contained  double-sided detectors to measure also $z$ coordinates.
The polar angle coverage of the VD was from
$25^\circ$ to $155^\circ$. Mini-strips and pixel detectors 
making up the Very Forward Tracker (VFT) have been added to the ends of the 
VD increasing the angular acceptance to include the regions from $10^\circ$ to 
$25^\circ$ and from $155^\circ$ to $170^\circ$~\cite{vft}.
The ID, covering polar angles between $15^\circ$ and $165^\circ$, was 
composed of a cylindrical drift chamber (inner radius 12\,cm and outer 
radius 22\,cm) surrounded by 5 layers of straw drift tubes with inner 
radius of 23\,cm and outer radius of 28\,cm. 
The TPC, the principal tracking device of 
DELPHI, was a cylinder of 30\,cm inner
radius, 122\,cm outer radius and length 2.7\,m. Each 
end-plate was divided 
into 6 sectors, with 192 sense wires 
and 16 circular pad rows  used for 3 dimensional 
space-point reconstruction.
The OD consisted of
5 layers of drift cells at radii
between 198\,cm and 206\,cm, covering polar angles between $43^\circ$
and $137^\circ$.

The average momentum resolution for the charged particles in hadronic final
states was in the range 
$\Delta p/p^2 \simeq 0.001$ to 
0.01 (GeV/c)$^{-1}$, depending on which detectors
were used in the track fit~\cite{delphi}.

The electromagnetic calorimeters were the High density Projection
Chamber (HPC) covering the barrel region of
$40^\circ<\theta<140^\circ$, the Forward ElectroMagnetic Calorimeter
(FEMC) covering $11^\circ<\theta<36^\circ$ and
$144^\circ<\theta<169^\circ$, and the STIC, a Scintillator TIle
Calorimeter which extended coverage down to 1.66$^\circ$ from the beam
axis in either direction.  The 40$^\circ$ taggers were a series of
single layer scintillator-lead counters used to veto electromagnetic
particles that would otherwise have been missed in the region between
the HPC and FEMC.  A similar set of taggers was arranged at 90$^\circ$ to 
cover the gap between the two halves of the HPC. 
The efficiency to register a photon with energy
above 5\,GeV, measured with the LEP1 data, was above 99$\%$.  The hadron
calorimeter (HCAL) covered 98$\%$ of the solid angle. Muons with
momenta above 2\,GeV/c penetrated the HCAL and were recorded in a set of
muon drift chambers.

\section{Data Samples}
The data collected by the DELPHI detector during 1997 at 
$\sqrt{s}\simeq 183$\,GeV and 1998 at $\sqrt{s}\simeq 189$\,GeV,  
corresponding to integrated luminosities of 53\,pb$^{-1}$ and 158\,pb$^{-1}$ 
respectively, were analysed.

To evaluate background contaminations, different contributions from the 
Standard Model processes were considered. The background processes 
WW, W$e\nu_e$, ZZ, Z$e^+e^-$ and 
Z$/\gamma\rightarrow q\bar{q}(\gamma)$  were generated using 
PYTHIA~\cite{phytia}, 
while the events Z$/\gamma\rightarrow\tau^+\tau^-(\gamma), 
\mu^+\mu^-(\gamma)$ 
were produced by KORALZ~\cite{koralz} and DYMU3~\cite{dymu3} respectively. 
A cross-check was performed using the four-fermion final states generated 
with EXCALIBUR~\cite{excalibur}. The generator BABAMC~\cite{babamc}  
was used for the Bhabha scattering. Two-photon interactions leading to 
leptonic and hadronic final states were produced by the BDK~\cite{bdk} and 
TWOGAM~\cite{twogan} programs, respectively. All the background events were 
passed through a detailed detector response simulation (DELSIM)   
and reconstructed as the real data~\cite{delphi}.

\begin{figure}
\begin{center}
\begin{tabular}{cc}
\epsfysize=7.5cm\epsffile{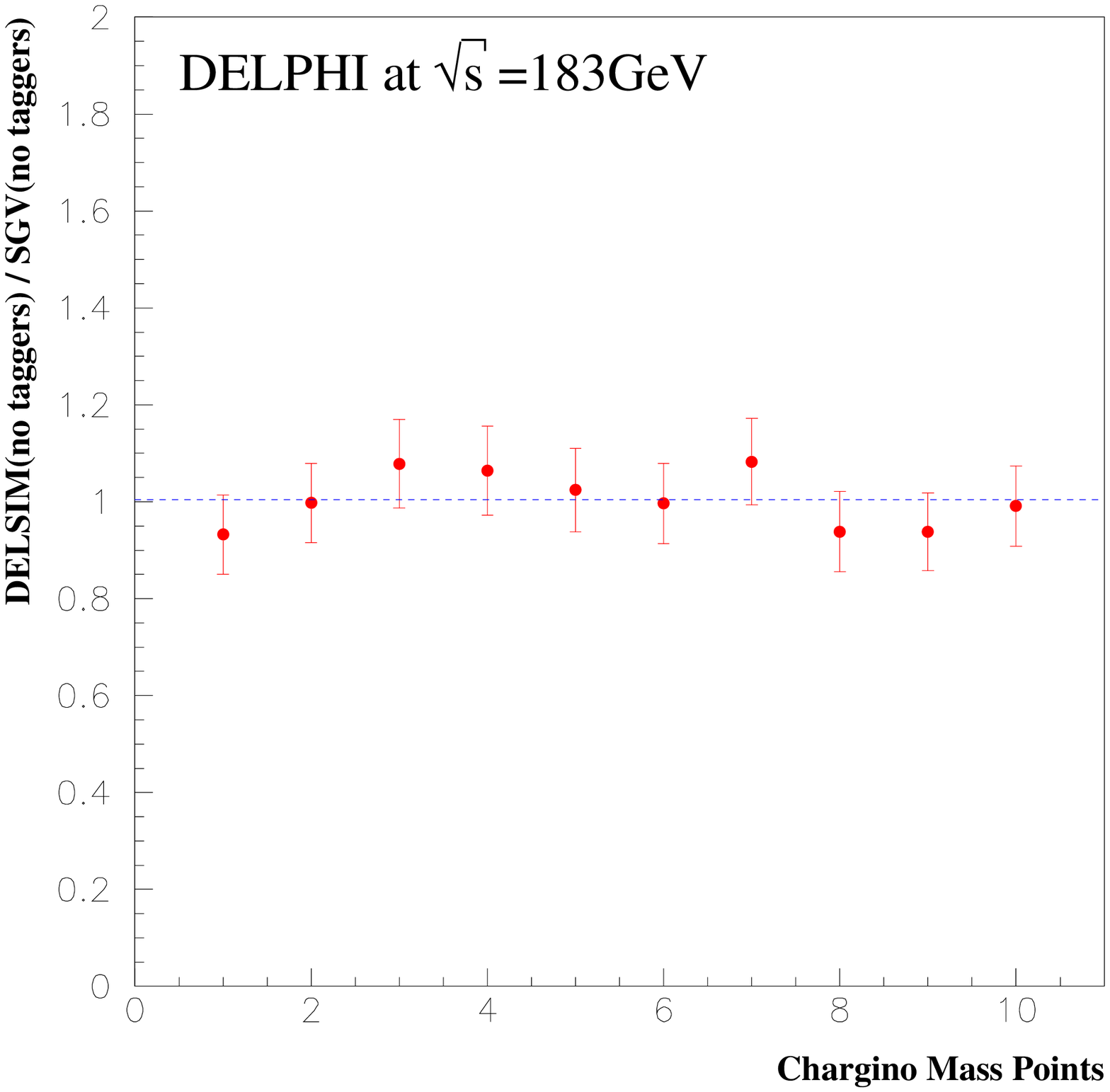} &  
\epsfysize=7.5cm\epsffile{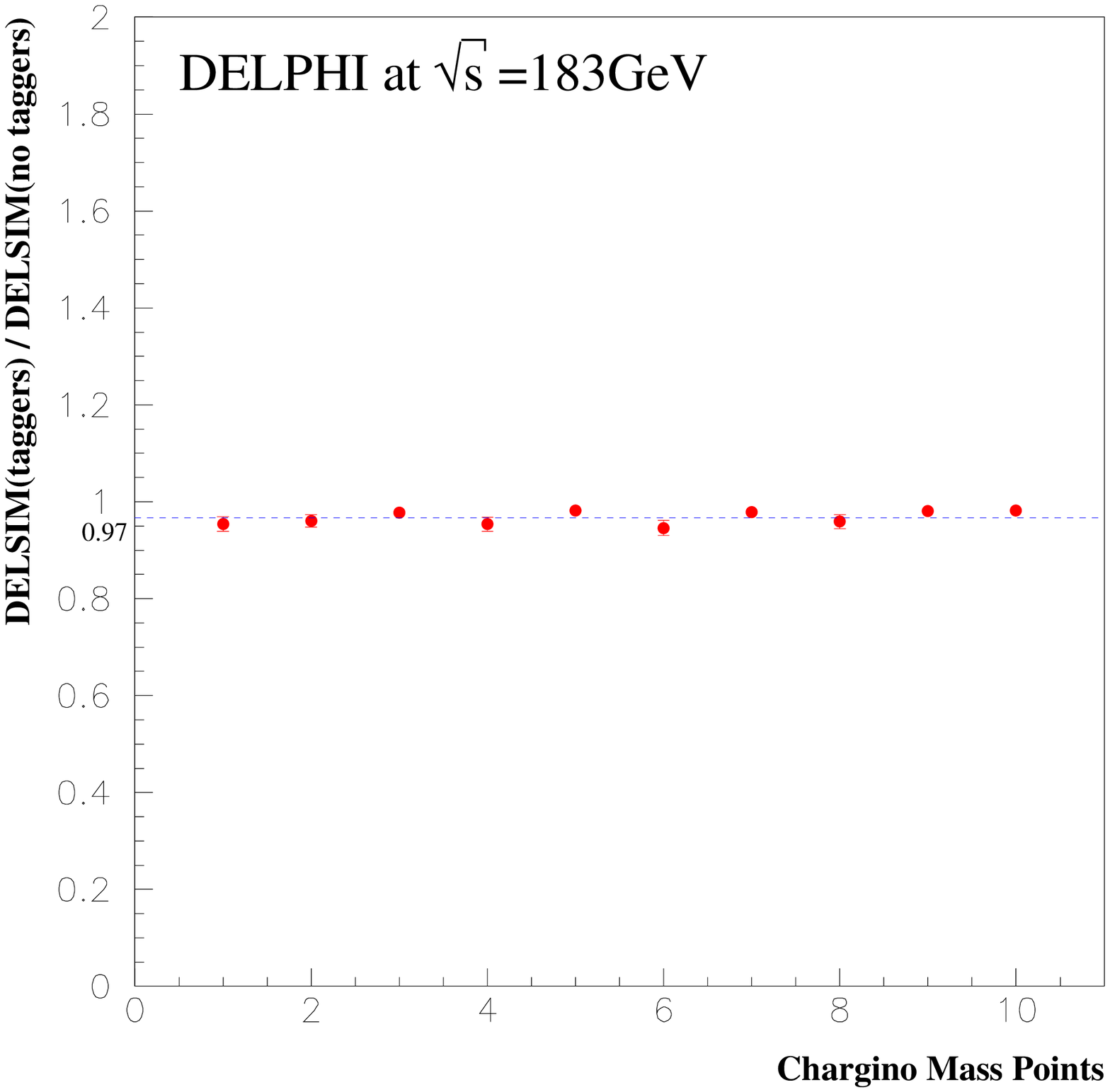} \\
(a) & (b) \\ 
\end{tabular}
\protect\caption{Efficiency correction factors for SGV simulated signals. 
(a) Selection efficiency ratio between the DELSIM simulated events and the 
SGV simulated events, if the taggers are not considered in the 
DELSIM simulated events. (b) Ratio between the 
selection efficiencies for the DELSIM simulated events with and without the 
tagger cut. The dashed line shows the average value for the efficiency 
correction factor that is equal to 1, if we compare DELSIM and SGV efficiencies 
(a) and equal to 0.97, if we use the tagger cut for the DELSIM simulated events 
(b).} 
\protect\label{factor}
\end{center}
\end{figure}

The program RP-generator II, described in reference~\cite{fernando}, was used  
to calculate the masses, production cross-sections and decay branching ratios of the chargino.
The chargino pair production was considered for different 
values of the $R$-parity violation parameter ($\epsilon$) and at several 
points of the MSSM parameter space ($\tan\beta, \mu, {\rm M}_2$).
For the signal, a faster simulation program  
SGV\footnote{The program ``Simulation a Grande Vitesse" (SGV) is described in http://delphiwww.cern.ch/\~{}berggren/sgv.html} was used to check the points that were not generated by the 
full DELPHI simulation program (DELSIM). The SGV program
does not simulate the DELPHI taggers. 
To correct for this effect, ten chargino mass points, with 1000 events each, 
were simulated by DELSIM and the selection efficiencies\footnote{The 
efficiency of the chargino selection is defined as the number of events 
satisfying the cuts defined in Secion~\ref{search_183} divided by the total number of generated chargino
events.} 
calculated from the two simulations were compared. Any differences between 
the two were used as a correction factor, shown in figure~\ref{factor}. 

\section{Chargino Searches}
\label{search}
With the $R$-parity spontaneous breaking, the chargino can decay 
through an $R$-parity conserving vertex into $\tau^\pm J$ events. 
Due to the undetectable Majoron, such events have the topology of two 
taus acoplanar with the beam axis plus missing energy. 
To select events with this signature it was required that 
the charged and neutral particles were well reconstructed and that the  
total momentum transverse to the beam was greater than 4\,GeV/c. 
A particle was considered as well reconstructed if it had a momentum 
between 1\,GeV/c and the beam momentum 
and a polar angle between 30$^\circ$ and 150$^\circ$. 
As a result 4006 events at 183\,GeV and 11350 at 189\,GeV were selected. 
The efficiency of detecting signal events at this level of selection was 
around 47\,\%. The simulated remaining backgrounds 
are detailed in table~\ref{data}.
 
Events were also required to have less than 7 charged particles and 
no signal in the 40$^\circ$ or 90$^\circ$ taggers. It was further required 
that the events consisted of 
two clusters of charged and neutral particles, 
each cluster with invariant mass below 5.5\,GeV/c$^2$ and with 
an acoplanarity\footnote{The acoplanarity is defined as the complement 
of the angle between the clusters when projected onto the plane 
perpendicular to the beam.} between 5$^\circ$ and 176$^\circ$. 
The clusters were constructed by considering all combinations of 
assigning the charged particles in the event into two groups. Neutral 
particles were then added to the groups such that the mass remains below the 
cut value and a neutral that can not be added to either of the two groups 
is considered as isolated.  

Events with forward going secondaries were avoided by rejecting any with 
energy measured in a 30$^\circ$ cone around the beam axis. 
Some of the energy in the forward cone resulted from noise and other 
backgrounds which were not included in the simulation of the signal. It was 
estimated that $\sim$ 20\% of any signal would be rejected by this 
selection and the efficiency was appropriately corrected. 
This preliminary selection resulted in 152 observed events at 183\,GeV and 
415 at 189\,GeV, as shown in table~\ref{data}, with a typical   
efficiency for the signal of $\sim$37\,\%.

\subsection{Event Selection at 183\,GeV}
\label{search_183}
To reject the 
radiative return to the Z background, no events with isolated photons with more 
than 5\,GeV were accepted. The $\gamma \gamma$ and $\mu^+\mu^-(\gamma)$ backgrounds
were reduced by requiring that the events had at least one charged particle with
momentum between 5\,GeV/c and 60\,GeV/c. To reduce the $\tau^+\tau^-(\gamma)$ background
the square of transverse momentum with respect to the thrust axis divided by
the thrust had to be above 0.75\,(GeV/c)$^2$.

To reduce the $\gamma \gamma$ background further, events with  momentum of 
their most energetic charged particle ($P_{max}$) below 10\,GeV/c had to have 
 total momentum transverse to the beam above 10.5\,GeV/c.
For events with $P_{max}>10$\,GeV/c, the main remaining contamination
comes from Z$/\gamma\rightarrow\tau^+\tau^-$ and WW. For those, if the 
acoplanarity was below 15$^\circ$, the angle between the missing momentum 
and the beam had to be greater than $30^\circ$.
On the other hand, if the acoplanarity was above 15$^\circ$, it was required
that the momentum of the most energetic particle was below 23.5\,GeV/c and
the angle between the missing momentum and the beam was greater 
than $34.5^\circ$.
 
\begin{figure}
\begin{center}
\begin{tabular}{cc}
\epsfysize=7.cm\epsffile{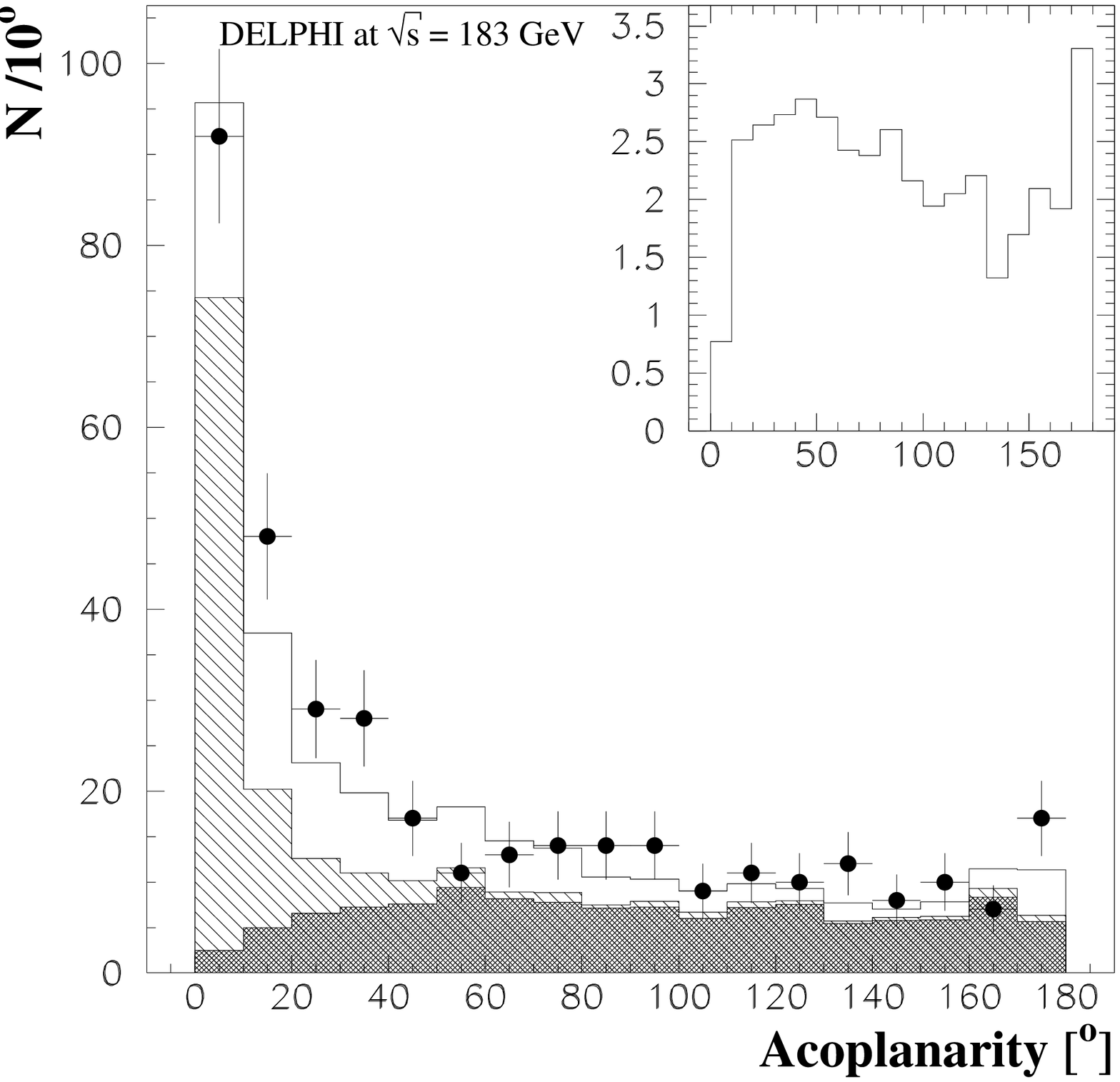} & 
\epsfysize=7.cm\epsffile{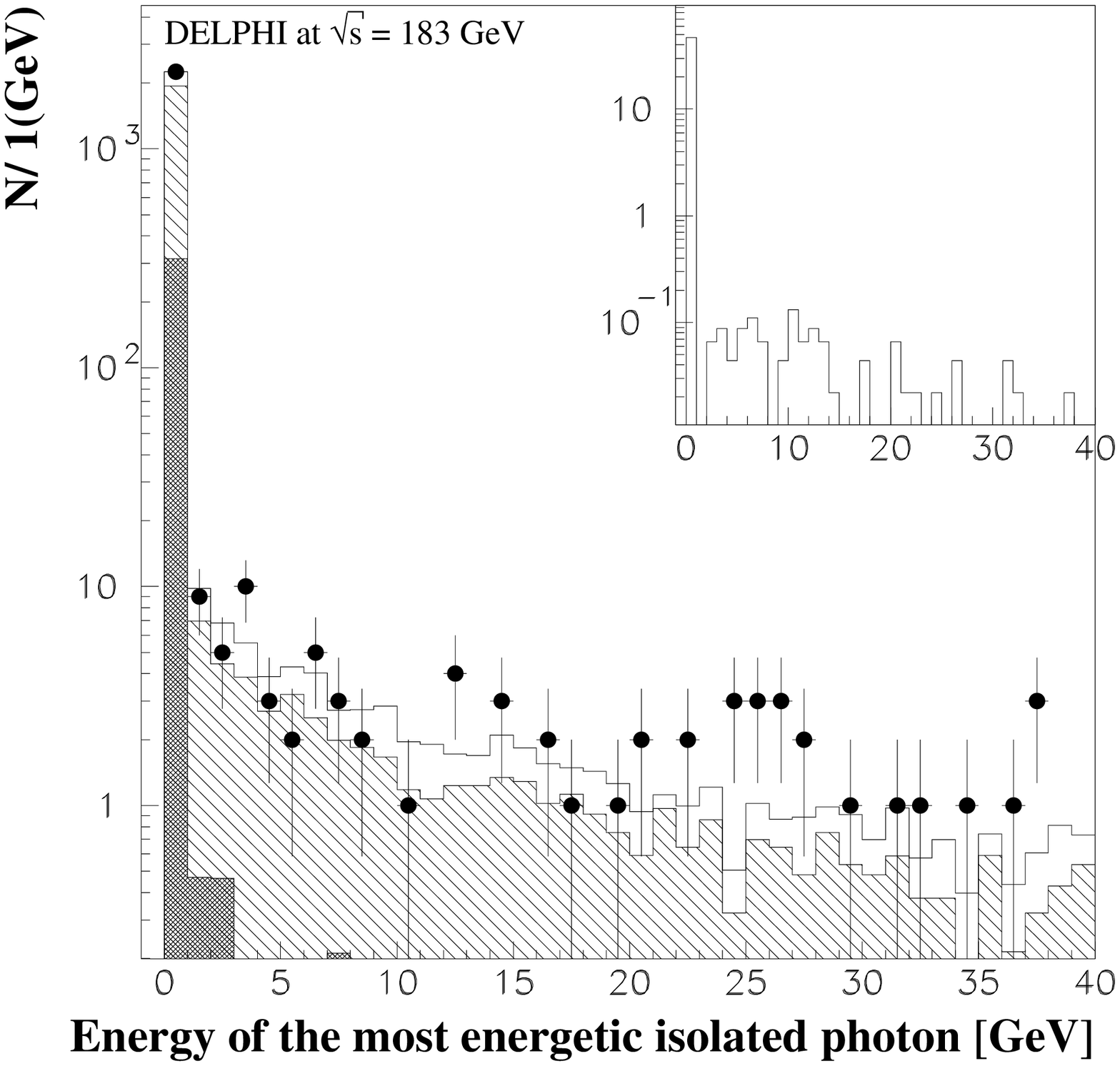} \\ 
(a) & (b) \\
\epsfysize=7.cm\epsffile{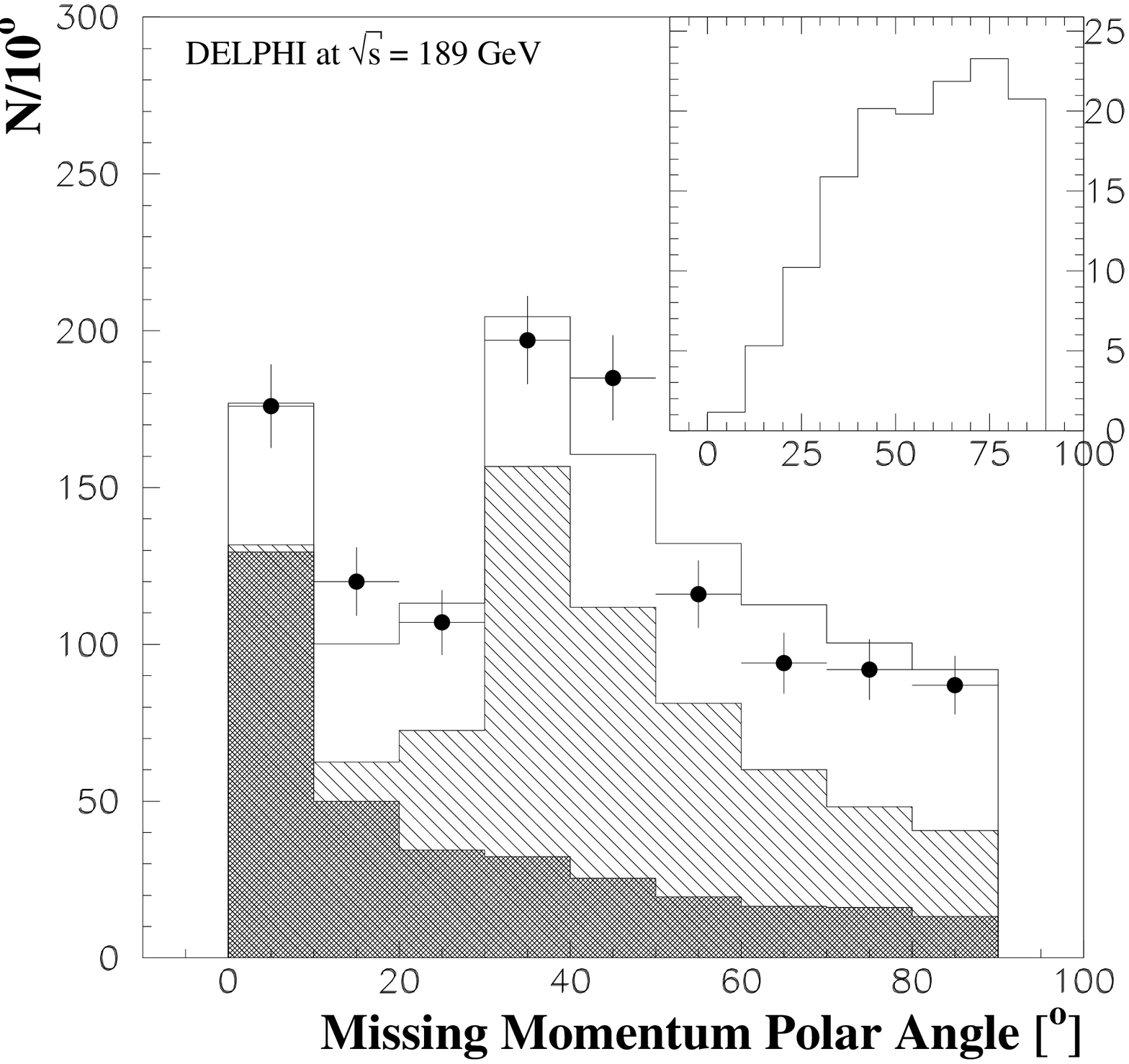}  &
\epsfysize=7.cm\epsffile{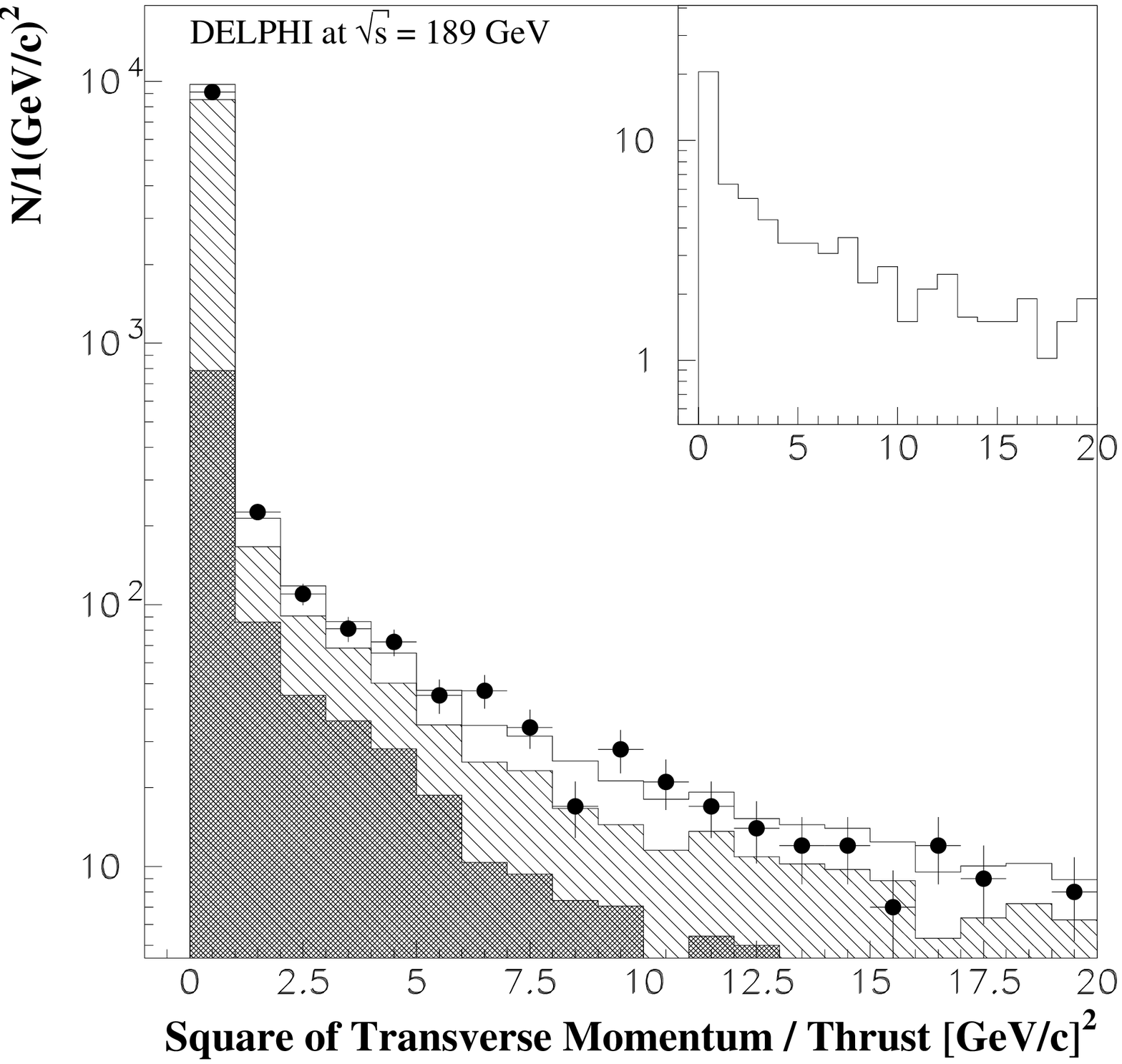} \\ 
(c) & (d) \\
\end{tabular}
\protect\caption{Distribution of (a) acoplanarity, (b) energy of the most 
energetic isolated photon, (c) angle between the missing momentum 
and the beam-axis and (d) square of transverse momentum 
with respect to the thrust axis divided by the thrust, requiring 
that the events had two clusters of well reconstructed neutral and 
charged particles, less than 7 charged particles and a total transverse 
momentum above 4\,GeV/c. 
For the acoplanarity and the missing momentum polar angle distributions
it was also required that the square 
of the transverse momentum with respect to the thrust axis divided by the 
thrust was above 0.75\,(GeV/c)$^2$ and above 1\,(GeV/c)$^2$, respectively.
The points with error bars show 
the real data, while the white histograms show the total simulated background. 
The distributions corresponding to the $\gamma\gamma$ background and the Bhabha 
scattering are shown as dark and hatched histograms, respectively. An example 
of the two body decay mode $\tilde{\chi}^\pm\rightarrow \tau^\pm + J$ 
behaviour for $\tan\beta=2$,
$\mu=100$\,GeV/c$^2$ and M$_2=400$\,GeV/c$^2$ is shown in the inserts for 
each plot.} 
\protect\label{data_183_189}
\end{center}
\end{figure}

\begin{figure}
\begin{center}
\begin{tabular}{cc}
\epsfysize=7.5cm\epsffile{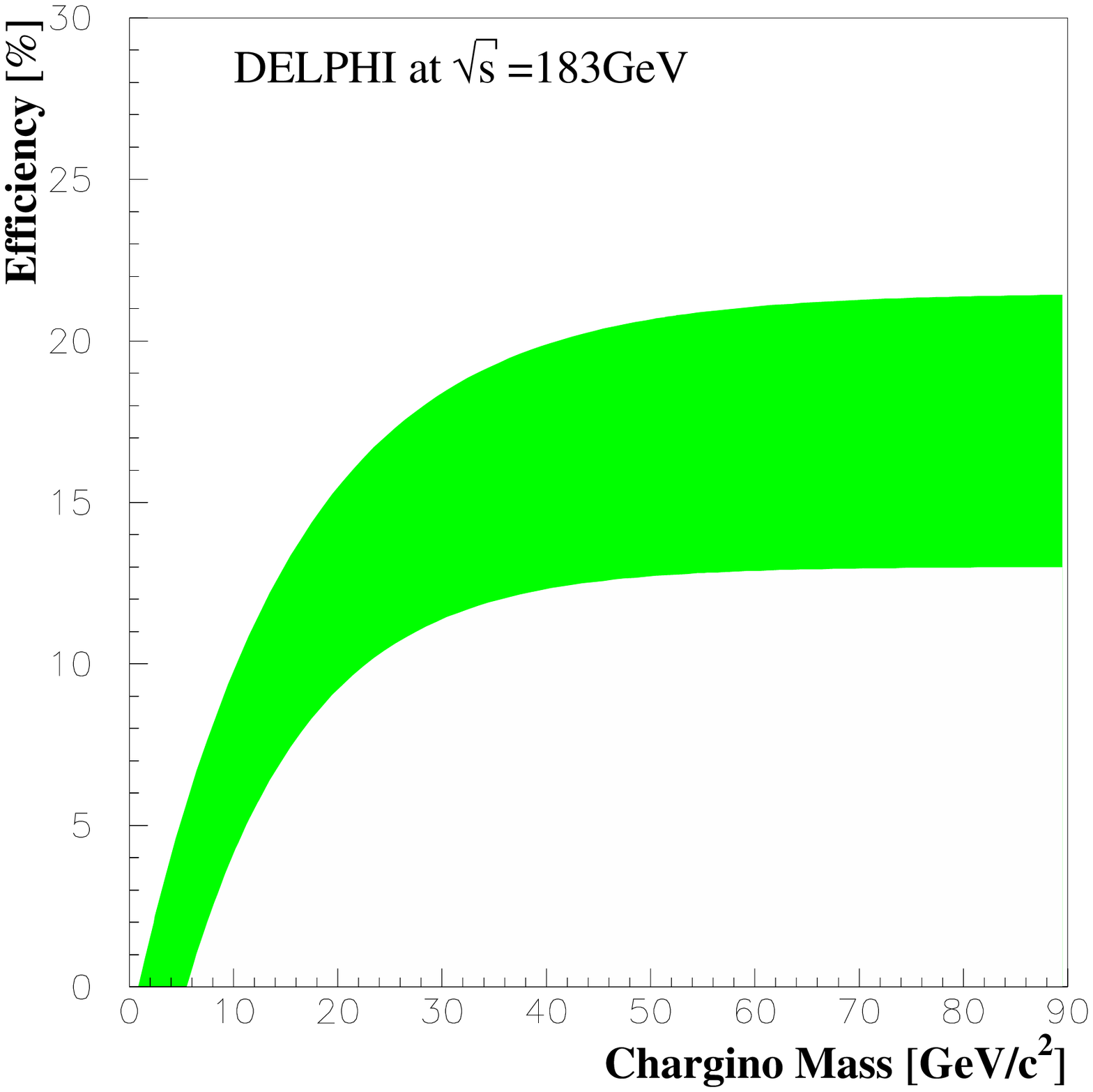} &
\epsfysize=7.5cm\epsffile{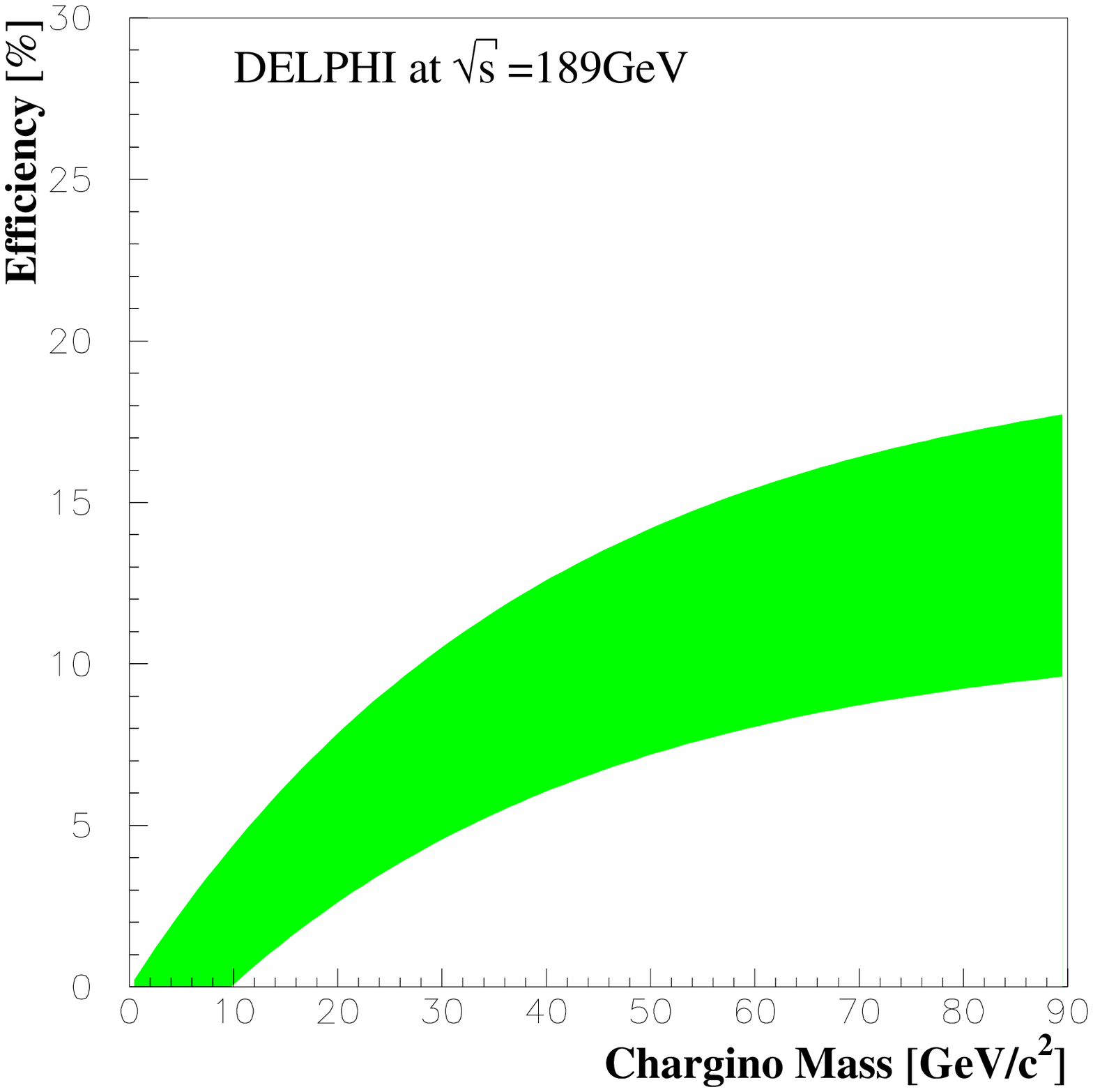}  \\  
(a) & (b) \\
\end{tabular}
\protect\caption{Chargino detection efficiency as a function of the chargino 
mass for (a) $\sqrt{s}=183$\,GeV and (b) $\sqrt{s}=189$\,GeV,  
considering only the two body decay mode $\tilde{\chi}^\pm\rightarrow \tau^\pm + J$. 
The bands correspond to the statistical uncertainties 
combined with the effect of generating events with different MSSM 
parameters M$_2$ and $\mu$, which have been varied in the ranges 
$40\,\rm{GeV}/c^2 \le \mathrm{M}_2 \le 400\,\rm{GeV}/c^2$ and 
$-200\,\rm{GeV}/c^2 \le \mu \le 200\,\rm{GeV}/c^2$, for $\tan\beta=2$.} 
\protect\label{eff}
\end{center}
\end{figure}

Figures~\ref{data_183_189}a and~\ref{data_183_189}b show the agreement between data and 
simulated background events after a preliminary selection, while 
figure~\ref{eff}a shows the dependence of the signal detection efficiency 
on the chargino mass. The selection criteria result in 6 observed events 
detected with a signal detection efficiency of around 18\,\%.

\subsection{Event Selection at 189\,GeV}
Since LEP delivered a higher luminosity for this energy and the WW background 
increased, tighter cuts were applied.
The required acoplanarity had to be between 10$^\circ$ and 176$^\circ$ and
no events with an isolated photon were accepted. The momentum of each of the two
 particle clusters had to be above 5\,GeV/c and below 55\,GeV/c and
the square of transverse momentum with respect to the thrust axis divided by
the thrust had to be above 1.0\,(GeV/c)$^2$.
All the events had to have the angle between the missing momentum 
and the beam greater than $35^\circ$.

The $\gamma \gamma$ background was mainly reduced by requiring a total momentum transverse to the beam greater than 9\,GeV/c. Events from WW processes were 
reduced by requiring that the momentum of the most energetic particle was below 23\,GeV/c. 

If one cluster had a momentum above 10\,GeV/c and the acoplanarity was less 
than 15$^\circ$ it was also required that the value of the effective centre-of-mass 
energy after any initial state radiation ($\sqrt{s'}$)~\cite{sprime}
 did not fall in 
the region between 90\,GeV and 94\,GeV. For an acoplanarity above  
15$^\circ$, the angle between the missing momentum
and the beam was required to be greater than $40^\circ$ 
and the visible mass lower than 
70\,GeV/c$^2$. 

Figure~\ref{data_183_189}c and~\ref{data_183_189}d show the agreement 
between data and simulated background events after a preliminary selection, 
 while figure~\ref{eff}b shows the dependence of
the signal detection efficiency on the chargino mass. The selection
criteria result in 9 observed events with a signal detection efficiency of 
around 14\,\%.
 
\section{Results}

As a result of the selection procedure, 6 candidates of 
$\tilde{\chi}^\pm\rightarrow \tau^\pm + J$ were selected at 183\,GeV, 
with a background estimation of $6.3\pm0.4$ and a signal detection efficiency 
of 18\,\%. 
At 189\,GeV, 9 candidates were found, with an expected background of 
$9.6\pm0.4$ and a signal detection efficiency of 14\,\%. 
Table~\ref{data} summarises the number of accepted events in the data, 
together with the predicted number of events from background sources. 
The systematic and statistical errors
on the simulated background calculation are insignificant compared to the
experimental statistical accuracy.

\begin{figure} 
\begin{center}
\mbox{\epsfysize=15.cm\epsffile{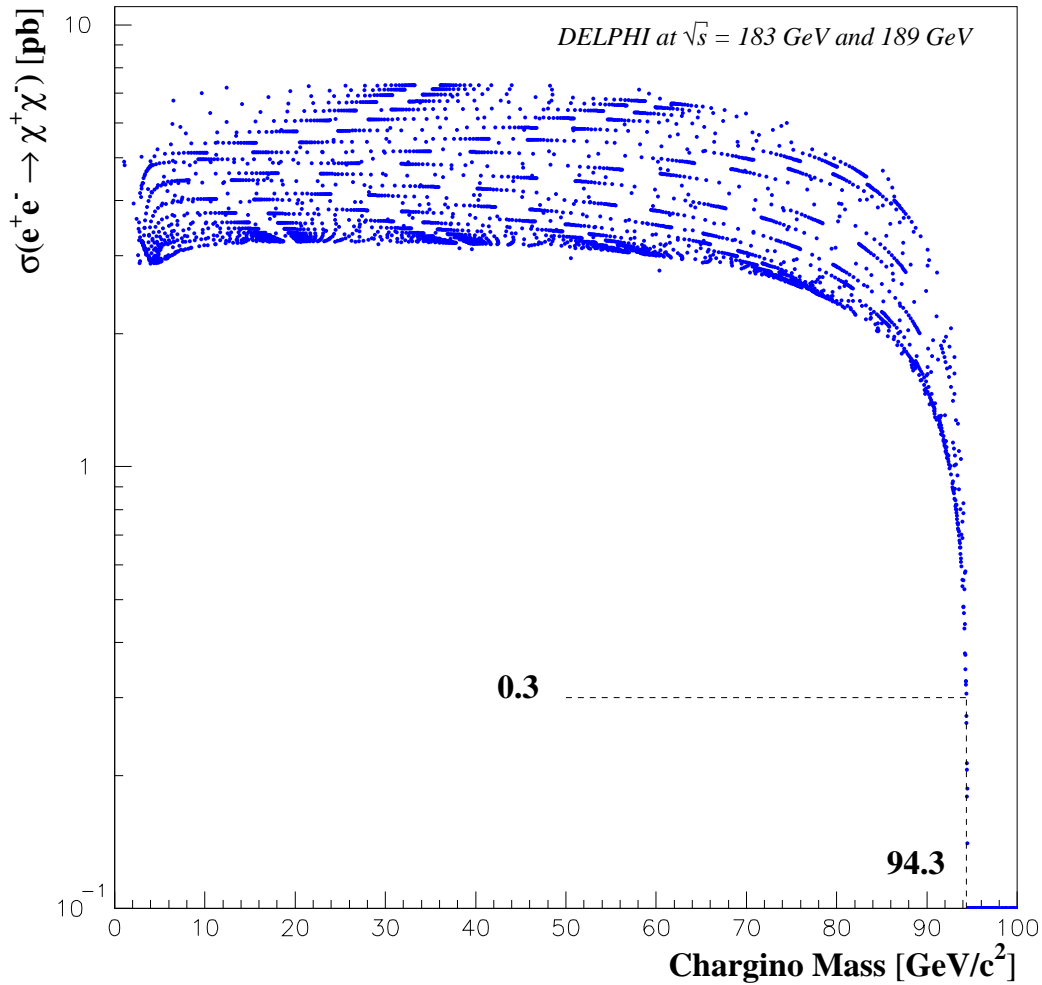}}
\protect\caption{Expected $e^+e-\rightarrow\tilde{\chi}^+\tilde{\chi}^-$ 
cross-section at 189\,GeV (dots) as a function of the chargino mass, 
assuming a heavy sneutrino ($M_{\tilde{\nu}}\geq 300$\,GeV/c$^2$). 
The dots correspond to the generated events at different chargino masses 
for the MSSM parameter ranges: 
$40\,\rm{GeV}/c^2 \le \mathrm{M}_2 \le 400\,\rm{GeV}/c^2$, 
$-200\,\rm{GeV}/c^2 \le \mu \le 200\,\rm{GeV}/c^2$ and $2\le \tan\beta\le 40$. 
At the 95\% confidence level, assuming the chargino decays mainly to 
$\tau^\pm J$, the maximum allowed chargino production cross-section in 
the excluded mass region is 0.3\,pb and the corresponding lower mass limit is 
94.3\,GeV/c$^2$.}
\protect\label{xsec}
\end{center}
\end{figure}

\begin{figure}
\begin{center}
\begin{tabular}{cc}
\epsfysize=7.5cm\epsffile{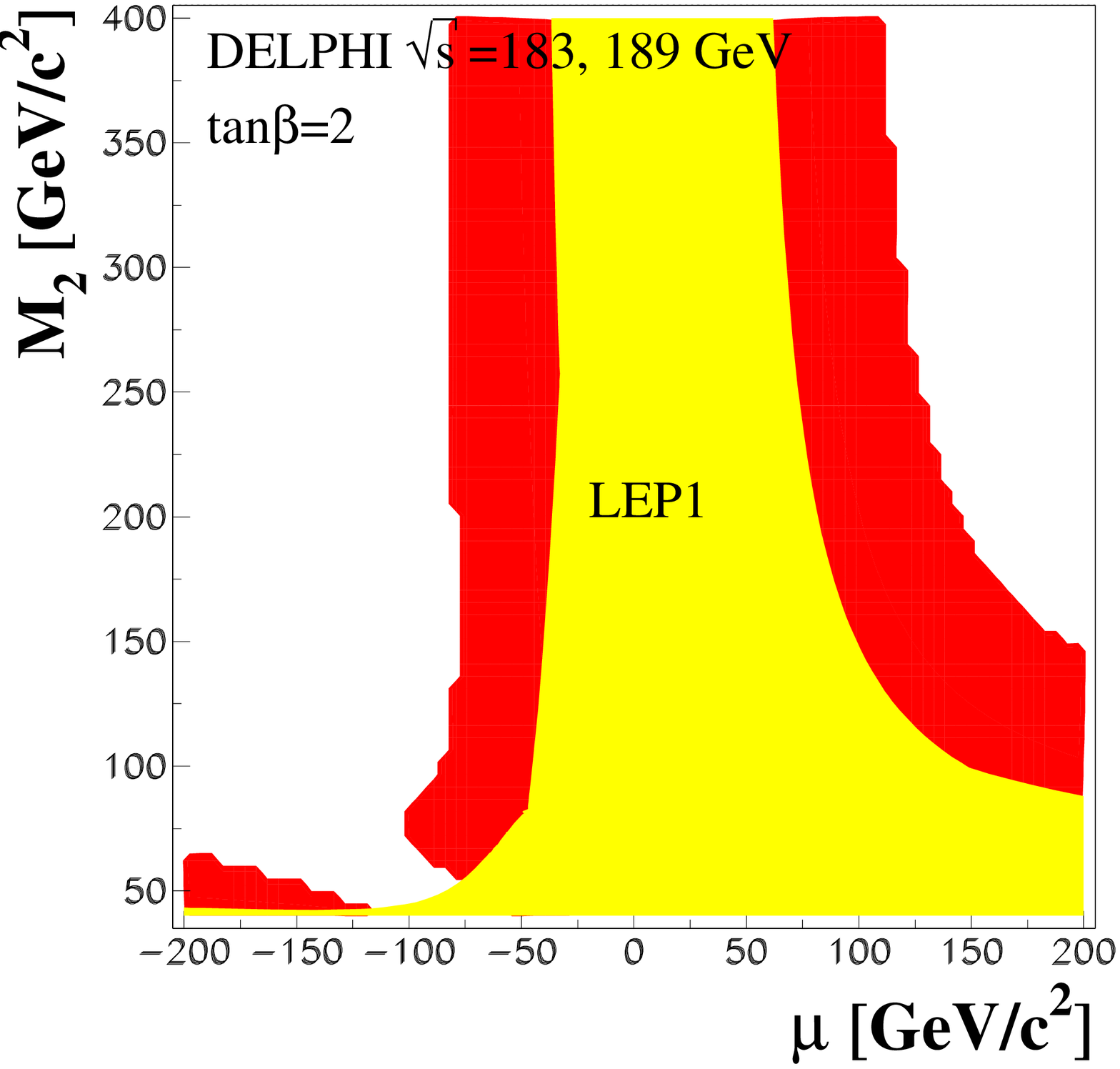} &
\epsfysize=7.5cm\epsffile{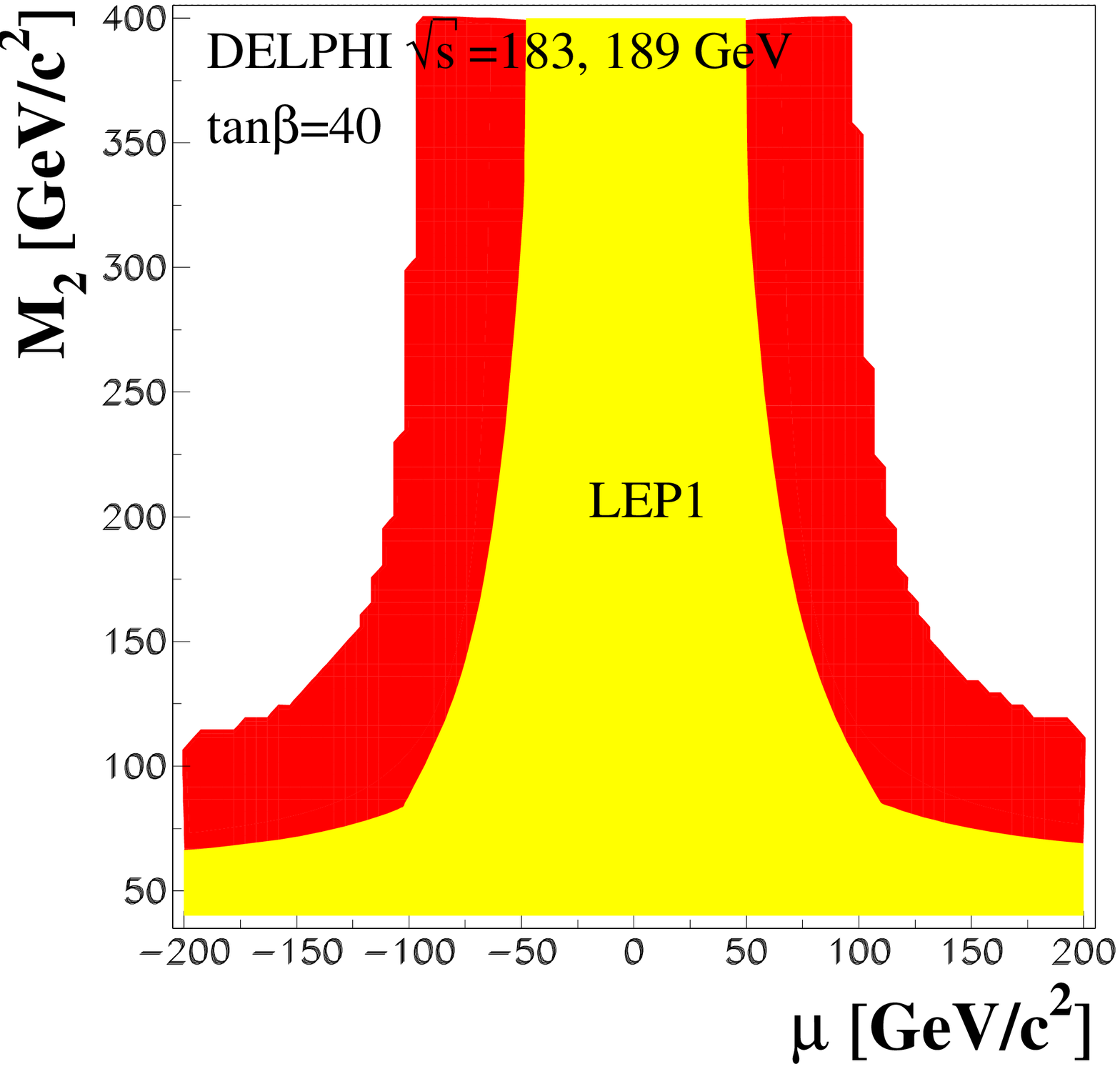} \\
(a) & (b) \\
\end{tabular}
\protect\caption{Regions in the $\mu$, $\rm{M}_2$ parameter space excluded at 
the 95\% confidence level for (a) $\tan\beta=2$ and (b) $\tan\beta=40$, assuming 
$M_{\tilde{\nu}}\geq 300$\,GeV/c$^2$. The exclusion area 
obtained with the $\tilde{\chi}^\pm \rightarrow \tau^\pm J$ search is shown in 
dark grey and the corresponding area excluded by the LEP1 data\cite{LEPI} 
is shown in light grey. The hole seen on plot (a) around M$_2=50$\,GeV/c$^2$ 
and $\mu=-120$\,GeV/c$^2$ is due to the low branching ratio (below 5\%) for the  
$\tilde{\chi}^\pm \rightarrow \tau^\pm J$.} 
\protect\label{exclusion}
\end{center}
\end{figure}

Assuming the chargino decays exclusively to $\tau^\pm J$, the data at 
183\,GeV and 189\,GeV, mentioned in the previous paragraph, 
were combined and the standard procedure described in~\cite{pdg} 
was used to obtain a 95\% confidence level upper limit on the allowed 
cross-section and a corresponding lower limit on the chargino mass; 
both are shown in figure~\ref{xsec}. Although the signal detection efficiency 
varies inside a certain band, as shown in  Figure~\ref{eff}, the lower 
limit on the 
chargino mass is not sensitive to this variation. 
The excluded domains of the MSSM parameter space for $\tan\beta=2$ and 
$\tan\beta=40$ are shown in figure~\ref{exclusion}. The limit obtained 
with this $\tilde{\chi}^\pm \rightarrow \tau^\pm J$ search 
substantially  extends the limit derived from the Z$^0$ line shape measured at 
LEP1~\cite{LEPI}.

\section{Conclusion}
Searches for spontaneous $R$-parity violating signals used 
a data sample of about 211\,pb$^{-1}$ collected by the DELPHI detector 
during 1997 and 1998 at centre-of-mass energies of 183\,GeV and 189\,GeV. 
In the present analysis it was assumed that the $R$-parity breaking occurs in 
the third generation and, as a consequence, the lightest chargino decays 
mainly through the two-body decay mode $\tilde{\chi}^\pm\rightarrow \tau^\pm + J$.
No evidence for $R$-parity spontaneously breaking has been observed, assuming a
sneutrino mass above 300\,GeV/c$^2$.  

In the search for $\tilde{\chi}^\pm\rightarrow \tau^\pm + J$, 15 candidates 
were selected, with  $15.9\pm0.6$ expected from SM processes.
This allowed an upper limit on the chargino production cross-section of 
0.3\,pb and a lower limit on the chargino mass of 94.3\,GeV/c$^2$ to be 
obtained at 95\% confidence level. The limit obtained with the present search 
substantially extends the general LEP1 limit~\cite{LEPI}.

\section{Acknowledgements}
\vskip 3 mm
 We are greatly indebted to our technical 
collaborators, to the members of the CERN-SL Division for the excellent 
performance of the LEP collider and to the funding agencies for their
support in building and operating the DELPHI detector.\\
We acknowledge in particular the support of \\
Austrian Federal Ministry of Science and Traffics, GZ 616.364/2-III/2a/98, \\
FNRS--FWO, Belgium,  \\
FINEP, CNPq, CAPES, FUJB and FAPERJ, Brazil, \\
Czech Ministry of Industry and Trade, GA CR 202/96/0450 and GA AVCR A1010521,\\
Danish Natural Research Council, \\
Commission of the European Communities (DG XII), \\
Direction des Sciences de la Mati$\grave{\mbox{\rm e}}$re, CEA, France, \\
Bundesministerium f$\ddot{\mbox{\rm u}}$r Bildung, Wissenschaft, Forschung 
und Technologie, Germany,\\
General Secretariat for Research and Technology, Greece, \\
National Science Foundation (NWO) and Foundation for Research on Matter (FOM),
The Netherlands, \\
Norwegian Research Council,  \\
State Committee for Scientific Research, Poland, 2P03B06015, 2P03B1116 and
SPUB/P03/178/98, \\
JNICT--Junta Nacional de Investiga\c{c}\~{a}o Cient\'{\i}fica 
e Tecnol$\acute{\mbox{\rm o}}$gica, Portugal, \\
Vedecka grantova agentura MS SR, Slovakia, Nr. 95/5195/134, \\
Ministry of Science and Technology of the Republic of Slovenia, \\
CICYT, Spain, AEN96--1661 and AEN96-1681,  \\
The Swedish Natural Science Research Council,      \\
Particle Physics and Astronomy Research Council, UK, \\
Department of Energy, USA, DE--FG02--94ER40817. \\

\pagebreak

\begin{table}
\begin{center}
\begin{tabular}{|l|r|r|} \hline

{\bf Centre-of-mass Energy}\hspace{3.3cm} & {~\bf 183\,GeV~} & {~~\bf 189\,GeV} \\\hline\hline 

{\bf Well reconstructed charged and neutral particles} & & \\ \hline

Observed events  & {\bf 4006} & {\bf 11350} \\ \hline 
Total Expected Background & {\bf 3769 $\pm$ 7} & {\bf 11511 $\pm$ 16} \\ \hline 
Bhabha scattering and Z$/\gamma\rightarrow ee,\mu\mu,\tau\tau,q\bar{q}$~~
& 3241 $\pm$ 4 & 
9858 $\pm$ 2 \\ \hline
4-fermion events except WW & 14 $\pm$ 1 & 82 $\pm$ 2 \\ \hline
$\gamma\gamma\rightarrow ee,\mu\mu,\tau\tau$ & 458 $\pm$ 5 
& 1393 $\pm$ 16 \\ \hline
W$^+$W$^-$ & 56 $\pm$ 1 & 178 $\pm$ 2 \\\hline\hline 

{\bf Preselection} & & \\ \hline

Observed events  & {\bf 152} & {\bf 415} \\ \hline 
Total Expected Background & {\bf 158 $\pm$ 4} & {\bf 494 $\pm$ 9} \\ \hline 
Bhabha scattering and Z$/\gamma\rightarrow ee,\mu\mu,\tau\tau,q\bar{q}$~~
& 65 $\pm$ 3 & 
207 $\pm$ 6 \\ \hline
4-fermion events except WW & 3 $\pm$ 1 & 10 $\pm$ 1 \\ \hline
$\gamma\gamma\rightarrow ee,\mu\mu,\tau\tau$ & 53 $\pm$ 2 & 158 
$\pm$ 7 \\ \hline
W$^+$W$^-$ & 37 $\pm$ 1 & 119 $\pm$ 2 \\\hline\hline 

{\bf Final Selection} & & \\ \hline

Observed events  & {\bf 6} &  {\bf 9} \\ \hline 
Total background & {\bf 6.3 $\pm$ 0.4} & {\bf 9.6 $\pm$ 0.4} \\ \hline
Bhabha scattering and Z$/\gamma\rightarrow ee,\mu\mu,\tau\tau,q\bar{q}$~~
& 1.0 $\pm$ 0.3 & 
0.6 $\pm$ 0.1 \\ \hline
4-fermion events except WW & 0.6 $\pm$ 0.1 & 1.2 $\pm$ 0.2 \\ \hline
$\gamma\gamma\rightarrow ee,\mu\mu,\tau\tau$ & 0.3 $\pm$ 0.1 & 0.2 
$\pm$ 0.2 \\ \hline
W$^+$W$^-$ & 4.4 $\pm$ 0.3 & 7.6 $\pm$ 0.3 \\\hline 

\end{tabular}
\end{center}
\protect\caption{Observed events (first row of each part) and the expected 
backgrounds (second to sixth row) at centre-of-mass 
energies of 183\,GeV and 189\,GeV. The preselection corresponds to 
the selection criteria described in the introduction of Section~\ref{search}. 
The errors quoted on the background 
correspond to the statistical uncertainties.} 
\protect\label{data}
\end{table}


\end{document}